\documentclass{nature}
\usepackage{graphicx}
\usepackage{amsmath}
\usepackage{color}
\usepackage{physics}
\usepackage{hyperref}
\usepackage{color}
\usepackage{soul}
\usepackage{slashed}
\usepackage[labelformat=empty]{caption}

\usepackage{amsfonts}
\usepackage[version=3]{mhchem} 

\makeatletter
\let\saved@includegraphics\includegraphics

\title{Cationic vacancies as defects in honeycomb lattices with modular symmetries}

\author{Godwill Mbiti Kanyolo$^1$ \& Titus Masese$^2$$^{,3}$}

\begin{document}

\maketitle

\begin{affiliations}
 \item Department of Engineering Science, The University of Electro-Communications 1-5-1 Chofugaoka, Chofu, Tokyo 182-8585, Japan
 
 \item Research Institute of Electrochemical Energy (RIECEN), National Institute of Advanced Industrial Science and Technology (AIST), 1-8-31 Midorigaoka, Ikeda, Osaka, 563-8577, Japan
 
 \item AIST-Kyoto University Chemical Energy Materials Open Innovation Laboratory (ChEM-OIL), Sakyo-ku, Kyoto 606-8501, Japan
 
\end{affiliations}

\begin{abstract}
Layered materials tend to exhibit intriguing crystalline symmetries and topological characteristics based on their two dimensional (2D) geometries and defects. We consider the diffusion dynamics of positively charged ions (cations) localized in honeycomb lattices within layered materials when an external electric field, non-trivial topologies, curvatures and cationic vacancies are present. The unit (primitive) cell of the honeycomb lattice is characterized by two generators, $J_1, J_2 \in \rm SL_2(\mathbb{Z})$ 
of modular symmetries in the special linear group with integer entries, corresponding to discrete re-scaling and rotations respectively. Moreover, applying a 2D conformal metric in an idealized model, we can consistently treat cationic vacancies as topological defects in an emergent manifold. The framework can be utilized to elucidate the molecular dynamics of the cations in exemplar honeycomb layered frameworks and the role of quantum geometry and topological defects not only in the diffusion process such as prediction of conductance peaks during cationic (de-)intercalation process, but also pseudo-spin and pseudo-magnetic field degrees of freedom on the cationic honeycomb lattice responsible for bilayers.
\end{abstract}

\newpage

\section{Introduction}

Layered materials tend to exhibit a myriad of intriguing crystalline symmetries and topological characteristics based on their two-dimensional (2D) geometries and defects.\cite{du2021engineering, kanyolo2021honeycomb} Since the discovery of graphene-based systems\cite{allen2010honeycomb} and layered materials\cite{du2021engineering} such as honeycomb layered materials\cite{kanyolo2021honeycomb}, a great deal of experimental and theoretical studies has been dedicated to illuminating the role the honeycomb lattice plays in the dynamics of electron quasi-particles and spin degrees of freedom, enriching our understanding of phenomena in materials ranging from high-temperature superconductors and 2D quantum hall systems to topological insulators and Kitaev materials.\cite{zhou2021high, von202040, kane2005z, kanyolo2021honeycomb} Despite their crystal-structural versatility and compositional tuneability attracting interest in various realms of solid-state (electro)chemistry, materials science, condensed matter physics and pioneering the discovery of next-generation energy storage materials\cite{kalantar2016two, kubota2020electrochemistry, kanyolo2021honeycomb, liu2019recent, he2012layered,schnelle2021magnetic, mcclelland2020muon}, theoretical and experimental studies centered on the honeycomb lattice have focused mainly on its effect on 2D electron and spin dynamics, thus rendering the behavior of larger particles such as positively-charged ions (cations) on the lattice understudied.

In particular, while the honeycomb lattice of graphene is formed by the carbon atoms, the electrodynamics studied is often centered around electron quasi-particles and spin degrees of freedom even when curvatures and topological defects are considered.\cite{allen2010honeycomb, mecklenburg2011spin, georgi2017tuning} By contrast, cations in layered materials can not only form the honeycomb lattice, but are also responsible for the electrodynamics during cation (de-)intercalation process.\cite{bera2020temperature, masese2018rechargeable} This poses a unique challenge to identifying the relevant quantum electrodynamics of cations in layered materials, not necessarily faced by other 2D systems. Intuitively (and in the continuum limit of the lattice), the quantum problem of electron dynamics in graphene with curvatures and defects is analogous to the problem of 1 + 2 dimensional quantum electrodynamics in a curved spaces, whereas the quantum problem of cations (charged) or their vacancies (uncharged) is analogous to the problem of 2D quantum gravity.\cite{kanyolo2021partition, gross1991two, holz1988geometry} 

\begin{figure}
\begin{center}
\includegraphics[width=\columnwidth,clip=true]{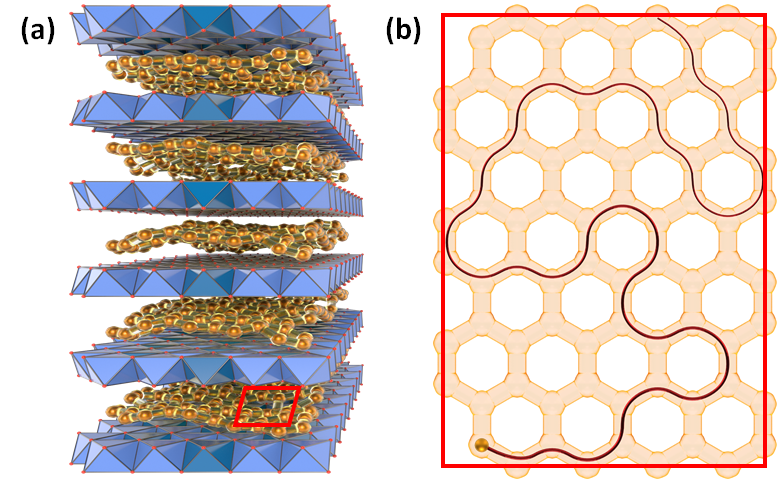}
 \caption{{\bf Figure 1:} (a) A schematic representation of the structure of exemplar honeycomb layered materials $A_4MD\rm O_6$, $A_3M_2D\rm O_6$ or $A_2M_2D\rm O_6$ wherein $A$ represents an alkali ion ($\rm Li, Na, K$, \textit{etc}.) or coinage metal ions such as Ag, whereas $M$ is mainly a transition metal species such as Co, Ni, Cu and Zn, and $D$ depicts a pnictogen or chalcogen metal species such as Sb, Bi and Te. The red rectangle at the base indicates the location of the schematic in {\bf Figure 1 (b)}; (b) Schematic depicting honeycomb-shaped diffusion pathways with vacant cationic sites in exemplar honeycomb layered materials. The maroon line shows a possible random diffusion pathway of the cation.}
\end{center}
\end{figure}

A specific class of honeycomb layered materials based on transition metal and semi-metal oxides has recently emerged adopting, \textit{inter alia}, chemical compositions embodied mainly by $A_4MD\rm O_6$, $A_3M_2D\rm O_6$ or $A_2M_2D\rm O_6$ wherein $A$ represents an alkali-ion ($\rm Li, Na, K$, \textit{etc}.) or coinage metal ions such as Ag, whereas $M$ is mainly a transition metal species such as Co, Ni, Cu, Zn, \textit{etc}. and $D$ depicts a pnictogen or chalcogen metal species such as Sb, Bi, As and Te.\cite{kumar2012novel, grundish2019electrochemical, nalbandyan2013crystal, skakle1997synthesis, smirnova2005subsolidus, politaev2010mixed, berthelot2012new, zvereva2012monoclinic, seibel2013structure, nagarajan2002new, zvereva2016orbitally, stratan2019synthesis, brown2019synthesis, uma2016synthesis, yadav2019new, zvereva2013new, roudebush2013structure, derakhshan2007electronic, viciu2007structure, evstigneeva2011new} These materials are often referred to as honeycomb layered oxides.\cite{kanyolo2021honeycomb} The structure of these honeycomb layered oxides is comprised of localized and de-localized $A$ cations sandwiched between slabs entailing $M$ atoms coordinated with oxygen around $D$ atoms, arranged in a honeycomb fashion (as shown in {\bf Figure 1(a)}). Thus, of specific interest to our work is the diffusion dynamics of the de-localized cations when an external electric field and non-trivial topology, curvatures and specific defects are present.\cite{kanyolo2020idealised, masese2021topologicalK2Ni2TeO6} 

Experimental investigations reveal the diffusion of cations to be largely restricted along honeycomb pathways in honeycomb layered tellurates such as $\rm Na_2 Ni_2TeO_6$ (as shown in {\bf Figure 1(b)}).\cite{bera2020temperature} While computational studies are consistent with this observation, they further suggest diffusion of cations in honeycomb pathways is restricted in honeycomb layered oxides for other materials as well, such as $\rm NaKNi_2TeO_6$ and $ A\rm_2Ni_2TeO_6$ where $A = \rm Na, K, Rb$ and $\rm Cs$ are cations with a large ionic radius, exhibiting a prismatic coordination with oxygen atoms of the $\rm Ni$ and $\rm Te$ octahedra forming the inter-layers.\cite{masese2018rechargeable, masese2021mixed, tada2022implications} In particular, Van der Waals forces and Coulomb repulsive forces tend to localize the cations in honeycomb lattices, creating a loosely-bound 2D non-Bravais hexagonal lattice with a two-cation basis known as the honeycomb lattice, which favors de-localization leaving vacancies in the hexagonal vertices only when sufficient activation energy from thermal fluctuations or the electric field, $\vec{E} = (E_x, E_y, 0)$ is present.\cite{wang2018na+, matsubara2020magnetism} Consequently, provided the ground state of the system devoid of activation energies was initially vacancy-free, the number of vacancies, $h$ in the honeycomb lattice is expected to closely correlate with the number of de-localized (mobile) cations, $\nu \simeq h \in \mathbb{N}$. Whence, the number of vacancies directly impacts the performance of the material as an effective cathode.\cite{hahn2013something} 

Meanwhile, in thin layers of superfluids, superconductors and liquid crystals deposited on curved 2D surfaces, topological defects are known to couple to 2D curvature degrees of freedom, leading to the identification of the number of topological defects as the Euler characteristic of the surface.\cite{musevic2006, mackintosh1991orientational, kamien2002geometry, vitelli2004anomalous, bowick2009two, turner2010vortices, mesarec2016effective} This lends credence to analogous treatments for cationic vacancies in layered materials.\cite{kanyolo2020idealised, kanyolo2021honeycomb} Moreover, layered materials demonstrating a bilayer arrangement of metal atoms (with each layer arranged in a triangular lattice) have been found, a vast majority being $\rm Ag$-based layered oxides and halides such as ${\rm Ag_2}M\rm O_2$ ($M = \rm Co, Cr, Ni, Cu, Fe, Mn, Rh$), $\rm Ag_2F$, $\rm Ag_6O_2$ (or equivalently as $\rm Ag_3O$), $\rm Ag_3Ni_2O_4$, and more recently ${\rm Ag_2}M_2\rm TeO_6$ (where $M = \rm Ni, Mg, Co, Cu, Zn$).\cite{allen2011electronic,schreyer2002synthesis,matsuda2012partially, ji2010orbital, yoshida2020static, yoshida2011novel, yoshida2008unique, yoshida2006spin, masese2021honeycomb, sorgel2007ag3ni2o4, argay1966redetermination, beesk1981x} Preliminary experimental and computational studies reveal that the bilayers represent a monolayer-bilayer phase transition of the honeycomb lattice, with the bifurcation mechanism not clearly understood.\cite{masese2021honeycomb, kanyolo2022conformal}

Herein, we consider how the honeycomb lattice of cations and emergent geometry constrains the model of cationic diffusion in such honeycomb layered oxides, leading to a rich topological description.\cite{kanyolo2020idealised} Consistently treating the number of vacancies, $h$ as the number of holes and handles (genus) of an emergent 2D manifold with a conformal metric, we conclude that the primitive basis and the corresponding Euler characteristic must obey the modular transformation\cite{cohen2017modular},
\begin{subequations}\label{modular_eq}
\begin{align}
    \chi (J\cdot k, h) = (\gamma k + \delta)^2\chi \left (k, h\right ),\\
    J \cdot k = \begin{pmatrix}
 \alpha & \beta \\
 \gamma & \delta \\
\end{pmatrix} \cdot k = \frac{\alpha k + \beta}{\gamma k + \delta},
\end{align}
\end{subequations}
invariant under the special linear group with integer entries, $J \in {\rm SL}_2$($\mathbb{Z}$) with $\alpha, \beta, \gamma, \delta \in \mathbb{Z}$ and $\det(J) = \alpha\delta - \beta\gamma = 1$, where $N = 2k$ is the total number of cation sites enclosed within the parallelogram with the primitive basis labeled by $dx$ and $dy$ as shown in {\bf Figure 2}.

\begin{figure}
\begin{center}
\includegraphics[width=\columnwidth,clip=true]{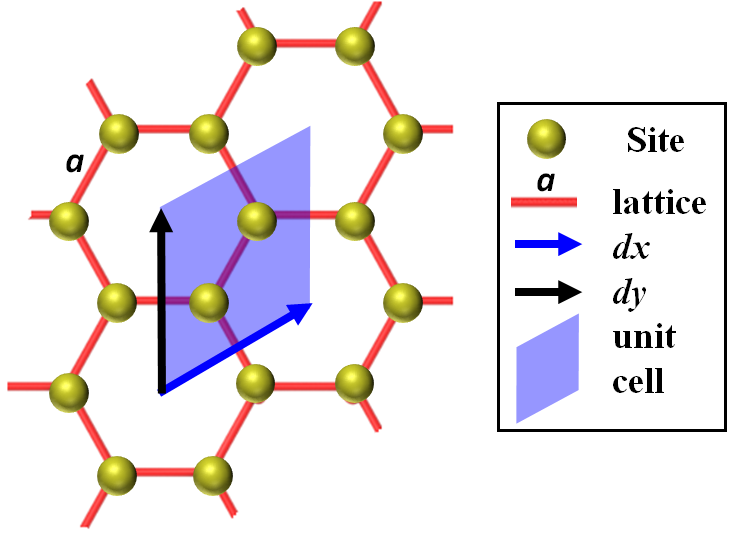}
 \caption{{\bf Figure 2:} Unit (primitive) cell ($dx/dy = N/2 = k = 1$) of localized cations arranged in a honeycomb lattice with lattice constant, $a$ showing the primitive vectors, $dx$ and $dy$. Every primitive cell is a parallelogram engulfing two (un-)occupied cationic sites, spanning partial interiors of four cationic-site hexagons.}
\end{center}
\end{figure}

In particular, the thermodynamics of the diffusive system of cations can be described by the partition function, $\mathcal{Z}$ of the form, 
\begin{subequations}\label{modular_partition_eq}
\begin{align}\label{partition_eq}
    \mathcal{Z} = \lim_{k \to \infty}\sum_{h \in \mathbb{N}} f_h\cosh\left (2\pi k\,\chi(k, h) \right ) \simeq \sum_{h \in \mathbb{N}}f_h\exp\left (2\pi k\,\chi(k, h) \right ),
\end{align}
which is invariant under the discrete re-scaling ($k \rightarrow k + 1 \simeq k$, for large $k$) and discrete rotations ($k \rightarrow -1/k$) of the primitive cell for small $k$, generated respectively by, 
\begin{align}\label{modular_symmetry_eq}
J = \left\{ J_1 = \begin{pmatrix}
 1 & 1 \\
 0 & 1 \\
 \end{pmatrix},\,\, J_2 = \begin{pmatrix}
 0 & -1\\
 1 & 0\\
 \end{pmatrix} \right\} \in {\rm SL}_2(\mathbb{Z}),
\end{align}
when the Euler characteristic can be written in a Fourier series,
\begin{align}\label{chi_eq}
    \chi(k, h) = b_n^{(0)} + b_n^{(2)} - \sum_{n = 1}^{h}b_n^{(1)}q^n(k),
\end{align}
\end{subequations}
$q(k) = \exp(2\pi i k)$, $b_n^{(p = 0, 1, 2)}$ is the $p$-th Betti number\cite{yano2016curvature} associated with the $n$-th primitive cell and topology labeled by $h$, $k = \bar{\beta}M/2 \in \mathbb{N}$ is an integer, $M$ is the average potential energy of the cations (Strictly speaking, it can be taken as the inverse of the Green's function of the system dominated by its potential energy, and hence can be complex-valued when the system has a finite life-time.\cite{mattuck1992guide}), $\bar{\beta} = 1/2\pi k_{\rm B}T$ is the `reduced' inverse temperature with $k_{\rm B}$ Boltzmann's constant and $f_h$ is a constant independent of $k$ but otherwise is dependent on the topology, $h$ and temperature, $T = 1/k_{\rm B}\bar{\beta}$. 

The conformal geometry and resultant Betti numbers can be taken to emerge from the underlying matrix group field theory of the self-interactions and quantum correlations of the cations in each primitive cell. In particular, a matrix large $\mathcal{N}$ group field theory can be considered, where $\mathcal{N} = \exp(2\pi k)$ is the size of the group.\cite{kanyolo2021partition} Consequently, this construction effectively treats cationic vacancies in honeycomb layered materials as topological defects in the underlying field theory, related to modular symmetries of the honeycomb lattice in the context of emergent 2D quantum geometries.\cite{gross1991two} For instance, we show that eq. (\ref{partition_eq}) follows from the partition function of $k$ pairs of cations forming the primitive cells of the honeycomb lattice, whereby each pair interacts via the Ising Hamiltonian due to emergent pseudo-spin and pseudo-magnetic degrees of freedom associated with the modular symmetry and broken conformal symmetry respectively. Analogous pseudo-degrees of freedom have also been considered in graphene-based systems.\cite{mecklenburg2011spin, georgi2017tuning} Thus, the framework can be utilized to elucidate the molecular dynamics of the cations in exemplar honeycomb layered frameworks and the role of quantum geometry and topological defects not only in the diffusion process such as prediction of conductance peaks during the cation (de-)intercalation process, but also pseudo-spin and pseudo-magnetic field degrees of freedom on the cationic honeycomb lattice whose interactions predict cationic bilayered frameworks.\cite{kanyolo2022conformal}

Hereafter, we shall set Planck's constant, the speed of electromagnetic waves in the material, $\bar{c}$, Boltzmann's constant, $k_{\rm B}$ and the elementary charge of the cations, $q_{\rm e}$ to unity, $\hbar = \bar{c} = k_{\rm B} = q_{\rm e} = 1$, and employ Einstein summation convention unless explicitly stated otherwise.

\section{The Model}

To build an intuitive geometric and topological picture of cationic diffusion, we begin by summarizing crucial results and clues from an idealized model, previously considered by the present authors in a separate publication.\cite{kanyolo2020idealised} Whilst the cations are positively charged, charge conservation requires the cationic vacancies created through de-localization by the electric field, $\vec{E}$ present in the 2D plane to be considered electrically neutral. Nonetheless, the vacancies can be treated as possessing a fictitious `magnetic moment' given by, 
\begin{subequations}\label{AC_eq}
\begin{align}
    \vec{\mu} = \bar{\beta}\vec{n},
\end{align}
where $\vec{n}$ is the unit normal vector to the 2D layer comprised of the honeycomb lattice of cations. Thus, the cations diffusing through and around the vacancies created by the electric field, $\vec{E}$ within the honeycomb lattice will introduce the `Aharonov-Casher' phase\cite{aharonov1984topological}, 
\begin{align}
    \Phi_{\rm AC}\Bigr|_{\partial \mathcal{A}} = \int_{\partial \mathcal{A}}(\vec{\mu}\times\vec{E})\cdot d\vec{x},
\end{align}
\end{subequations}
where $\partial \mathcal{A}$ is the boundary of a 2D patch, $\mathcal{A}$ of the manifold associated with the 2D layer spanning the primitive cells, where locally the Cartesian coordinates are given by $\vec{x} = (x, y, z)$. Aligning the manifold with the $x-y$ plane, the unit normal vector becomes $\vec{n} = (0, 0, 1)$. 

Applying Stokes' theorem and substituting respectively Maxwell's equations (Gauss' law) and the normalization,
\begin{subequations}\label{Gauss_law_eq}
\begin{align}
    \vec{\nabla}\cdot\vec{E}(x,y) = 4\pi\rho (x,y),\\
    \int^{\bar{\beta}/2}_{-\bar{\beta}/2}dz\int_{\mathcal{A}} dxdy\,\rho(x,y) = \nu,
\end{align}
\end{subequations}
where $\rho(x,y) = \rho_{\rm 2D}(x,y)/\bar{\beta}$, $\rho_{\rm 2D}(x,y)$ is the 2D cation number density, $\bar{\beta}$ is set as the integration cut-off scale in the $z$-direction and $\nu$ is the total number of mobile cations, yields the condition,
\begin{subequations}\label{Dirac_eq}
\begin{align}
    \Phi_{\rm AC}\Bigr|_{\partial \mathcal{A}} = 4\pi \nu, 
\end{align}
\end{subequations}
where $\nu \simeq h \in \mathbb{N} \geq 0$ is also the number of vacancies. 

Moreover, assuming the diffusion paths trace arc lengths defined by the 2D conformal metric,
\begin{subequations}\label{Liouville_eq}
\begin{align}
    ds^2 = g_{ab}dx^adx^b = \exp(2\Phi(x,y))(dx^2 + dy^2), 
\end{align}
where $g_{ab}$ is the 2D metric tensor, $\Phi(x,y)$ must satisfy Liouville's equation\cite{kanyolo2020idealised},
\begin{align}
    \nabla^2\Phi(x,y) = -K(x,y)\exp(2\Phi(x,y)),
\end{align}
\end{subequations}
with $K(x,y)$ the Gaussian curvature of the manifold. In fact, one can relate the two scalar functions, $\Phi$ and $\Phi_{\rm AC}$ by requiring that,
\begin{align}\label{set_eq}
    \Phi(x,y) = -\int \vec{n}\times\vec{\nabla}\Phi_{\rm AC} \cdot d\vec{x}.
\end{align}
Whence, by Stokes' theorem, the Euler characteristic of the manifold is given by the Gauss-Bonnet/Poincar\'{e}-Hopf theorem,
\begin{multline}\label{Gauss_Bonnet_eq}
    \chi = \frac{1}{2\pi}\int_{\mathcal{A}}K(x,y)\sqrt{\det(g_{ab})}\,dxdy
   = \frac{1}{2\pi}\int_{\mathcal{A}}K(x,y)\exp(2\Phi(x,y))\,dxdy\\
   -\frac{1}{2\pi}\int_{\mathcal{A}}\nabla^2\Phi(x,y)dxdy
   = -\frac{1}{2\pi}\int_{\mathcal{A}}\vec{n}\cdot\vec{\nabla}\times\vec{\nabla}\Phi_{\rm AC} dxdy,\\
   = -\frac{1}{2\pi}\int_{\partial\mathcal{A}}\vec{\nabla}\Phi_{\rm AC}\cdot d\vec{x}
   = -\frac{1}{2\pi}\Phi_{\rm AC}\Bigr|_{\partial \mathcal{A}} = -2\nu, 
\end{multline}
where we have used eq. (\ref{Dirac_eq}), eq. (\ref{Liouville_eq}) and eq. (\ref{set_eq}) to arrive at our result. 

For instance, for a compact orientable 2D manifold homeomorphic to $h$ number of simply-connected 2-tori, the Euler characteristic is given by, $\chi(h) = 2 - 2h$ where $h$ is the genus of the surface given by,
\begin{align}
    \nu = h - 1,
\end{align}
    which satisfies $\nu \simeq h$ for a large number of diffusing cations, $\nu \rightarrow \infty$. Thus, this avails the avenue to treat the number of cationic vacancies as the genus, $h$ which uniquely defines the emergent topology of the manifold. Moreover, using $F_{0i} = \vec{E} = (E_x, E_y, 0)$ and $\frac{1}{2}\varepsilon_{ijk}F_{jk} = \vec{B} = 0$ with $\varepsilon_{ijk}$ the 3D Levi-Civita symbol normalized as $\varepsilon_{123} = 1$, eq. (\ref{AC_eq}) and eq. (\ref{set_eq}) follow from the phase equations of motion\cite{kanyolo2020renormalization, kanyolo2020rescaling},
\begin{subequations}\label{phase_eq}
\begin{align}
    \partial_{\mu}\Phi =  \bar{\beta}\xi^{\nu}\eta_{\sigma\mu}\eta_{\rho\nu}F^{\sigma\rho},\\
    \partial_{\mu}\Phi_{\rm AC} = \bar{\beta} n^{\nu}\,\eta_{\sigma\mu}\eta_{\rho\nu}^*F^{\sigma\rho},
\end{align}
\end{subequations}
on the Minkowski metric, 
\begin{align*}
    ds_{\rm M}^2 = -\eta_{\sigma\rho}dx^{\sigma}dx^{\rho} = dt^2 - dx^2 - dy^2 - dz^2,
\end{align*}
where $\eta_{\mu\nu}$ is the Minkowski metric tensor, 
\begin{align}\label{Maxwell_eqs}
    \partial_{\mu} F^{\mu\nu} = 4\pi J^{\nu},\,\, \partial_{\mu}\!^*F^{\mu\nu} = 0,
\end{align}
are Maxwell's equations, $\xi^{\mu} = (1, \vec{0})$ and $n^{\mu} = (\vec{0}, 1)$ are time-like and space-like unit normal four-vectors respectively, $F_{\mu\nu} = \partial_{\mu}A_{\nu} - \partial_{\nu}A_{\mu}$ is the electromagnetic field strength, $A_{\mu}$ is the electromagnetic (U($1$)) gauge field, $^*F_{\mu\nu} = \frac{1}{2}\varepsilon_{\mu\nu\sigma\rho}F^{\sigma\rho}$ is the dual field strength with $\varepsilon_{\mu\nu\sigma\rho}$ the 4D Levi-Civita symbol normalized as $\varepsilon_{1234} = 1$ and $J^{\mu} = (\rho, \vec{J})$ is the current density of the cations. Thus, by eq. (\ref{phase_eq}), the 2D charge density is related to the Gaussian curvature by\cite{kanyolo2020idealised},
\begin{align}
    \rho_{\rm 2D}(x,y) = -\frac{1}{4\pi}K(x,y).
\end{align}

To incorporate diffusion in the formalism, we introduce the diffusion current given by,
\begin{subequations}\label{Diffusion_eq}
\begin{align}
    \vec{J}_{\rm AC} = -D\vec{\nabla}\rho = \rho_{\rm 2D}\,\vec{p},
\end{align}
corresponding to Fick's first law with $D$ the diffusion coefficient and $\vec{p}$ the center of mass momentum of the cations. Taking the cation number density to satisfy Boltzmann distribution at equilibrium,
\begin{align}\label{Boltzmann_eq}
    \rho(\Phi_{\rm AC}) \propto \exp\left (-\frac{1}{2}\bar{\beta}M\Phi_{\rm AC} \right ),
\end{align}
\end{subequations}
with $\frac{1}{2}M\Phi_{\rm AC}(x,y)$ a `gravitational' potential energy governing the diffusion dynamics and $M = 2/D$ a peculiarly defined center of mass effective mass using the diffusion coefficient,
and applying eq. (\ref{phase_eq}) yields,
\begin{subequations}\label{Langevin_eq}
\begin{align}
   \vec{p} = \vec{\nabla}\Phi_{\rm AC},\\
   0 = \frac{d\vec{p}}{dt} = -\bar{\beta}^{-1}\vec{p} + \vec{n}\times \vec{E}.
\end{align}
\end{subequations}
Thus, eq. (\ref{Langevin_eq}) correspond to the Hamilton-Jacobi equations for the cations with $\Phi_{\rm AC}$ corresponding to Hamilton's principal function, the second equation to the 2D Langevin equation\cite{lemons1997paul} and $\bar{\beta}$ to the mean-free time/path between collisions (friction term). Thus, this serves as the motivation for $\bar{\beta}$ appearing as the cut-off time and length scale in eq. (\ref{Gauss_law_eq}) and eq. (\ref{phase_eq}). 
Moreover, the peculiar relation, $M = 2/D$ can be better understood by applying the Virial theorem\cite{marc1985virial},
\begin{subequations}
\begin{align}\label{virial_eq}
     N/\bar{\beta} = \sum_{j = 1}^{N}\left \langle \frac{\vec{p}_j\cdot\vec{p}_j}{2\bar{m}} \right \rangle
     = \frac{1}{2}\sum_{j = 1}^{N}\left \langle \vec{r}_j\cdot\frac{\partial V(\vec{r}_j)}{\partial \vec{r}_j } \right \rangle
    = \sum_{j = 1}^{N}\langle V(\vec{r}_j) \rangle \equiv M,
\end{align}
where the averages are evaluated at equilibrium using,
\begin{align}
    \langle \cdots \rangle = \frac{\int (\cdots) \exp\left (-\bar{\beta}\mathcal{H}(\vec{p}_j,\vec{r}_j)  \right )\prod_{j = 1}^{N}d^2p_jd^2r_j}{\int \exp\left (-\bar{\beta}\mathcal{H}(\vec{p}_k,\vec{r}_k)  \right )\prod_{k = 1}^{N}d^2p_kd^2r_k}.
\end{align}
In this study, we shall consider the particular Hamiltonian for the cations,
\begin{align}
   \mathcal{H}(\vec{p}_j,\vec{r}_j) = \sum_{j = 0}^{N} \left (\frac{\vec{p}_j\cdot\vec{p}_j}{2\bar{m}}  + V(r_j) \right ),
\end{align}
\end{subequations}
with momenta, $\vec{p}_j$, displacement vectors, $\vec{r}_j$, $\bar{m} = 1/\bar{\beta}$ a mass per cation parameter defined as the inverse of the mean time/path between collisions, $\bar{\beta}$ and $V(r_{j}) \simeq \frac{1}{2}\bar{m}\mu^{-1}\sum_{k = 1}^{N}\vec{r}_k\cdot\vec{r}_j$ the leading interaction term in the potential energy defined proportional to $\mu$, the mobility of the cations. Typically, other terms such as the Vashishta-Rahma potential\cite{Vashishta1978}, which capture interactions of the cations with the slabs atoms especially oxygen, contribute higher order terms neglected herein. This requires that the diffusion coefficient, including cation-cation correlation terms\cite{vargas2020dynamic}, satisfy the Einstein-Smoluchowski relation, 
\begin{subequations}
\begin{align}
   D = \frac{1}{2\bar{\beta}}\left \langle \frac{1}{N}\sum_{j, k = 1}^N \vec{r}_j\cdot\vec{r}_k \right \rangle
   \simeq \frac{\mu}{2\pi N}\sum_{j = 1}^{N}\left \langle V(r_j) \right \rangle
   = \mu/\bar{\beta} = \mu M/\bar{\beta} M, 
\end{align}
\end{subequations}
as $\bar{\beta} \rightarrow \infty$, where we have used the result in eq. (\ref{virial_eq}). Thus, $D = 2/M$ requires we have $\mu M/2 = \bar{\beta}$.

Observe that, when cation-cation correlation ($j \neq k$) terms vanish, the diffusion coefficient becomes the self-diffusion coefficient, whereas $\sqrt{\mu}$ takes the role of frequency of the harmonic oscillator. Moreover, since the mobility is a constant, we can re-define it as $16\pi G \equiv \mu$, where $G \sim a^2$ and $a$ is taken to be the lattice constant with dimensions of $\rm length$. We thus have, $\bar{\beta} = 8\pi GM$ and $N = 2k = \bar{\beta}M = 4GM^2$, where $G$ is a gravitational constant, in obvious comparison with Schwarzschild black hole thermodynamics.\cite{kanyolo2022local}

\section{Results}

\subsection{Conductance Spikes:}

We note that, eq. (\ref{phase_eq}) and eq. (\ref{Diffusion_eq}) require that the diffusion current takes the Chern-Simons form\cite{dunne1999aspects},
\begin{subequations}
\begin{align}
    \vec{J}_{\rm AC} = \frac{k}{2\pi}\sigma\, (\vec{n}\times\vec{E}),
\end{align}
where $\sigma = 2\pi\bar{\beta} D\rho = \mu\rho$ is the conductivity of a single primitive cell, $\mu = 2\pi\bar{\beta} D$ is the mobility (Einstein-Smoluchowski equation) and,
\begin{align}\label{k_eq}
    k = \frac{N}{2} = \bar{\beta}M/2 \in \mathbb{N},
\end{align}
is the Chern-Simons level. However, the current (density) we are interested in is not necessarily the Hall current, $\vec{J}_{\rm AC}$ but the spatial part of $J^{\mu}$,
\begin{align}
    \vec{J} = \vec{n}\times\vec{J}_{\rm AC},
\end{align}
\end{subequations}
which couples to the electromagnetic tensor, $F_{\mu\nu}$ in Maxwell's equations given in eq. (\ref{Maxwell_eqs}). Consequently, this predicts conductance spikes whenever $k \rightarrow k + 1$ primitive cells containing $N = 2k \rightarrow N + 2 = 2k + 2$ cation sites are activated in the (de-)intercalation process. However, due to typically low measurement sensitivity in existing experimental data, the integer nature of the conductance cannot be ascertained. Nonetheless, distinct conductance spikes can be observed at low resolution (\textit{i.e} $k \rightarrow k + w$ with $w \in \mathbb{N} \gg 1$) at specific voltage values as shown {\bf Figure 3}.

\begin{figure}
\begin{center}
\includegraphics[width=\columnwidth,clip=true]{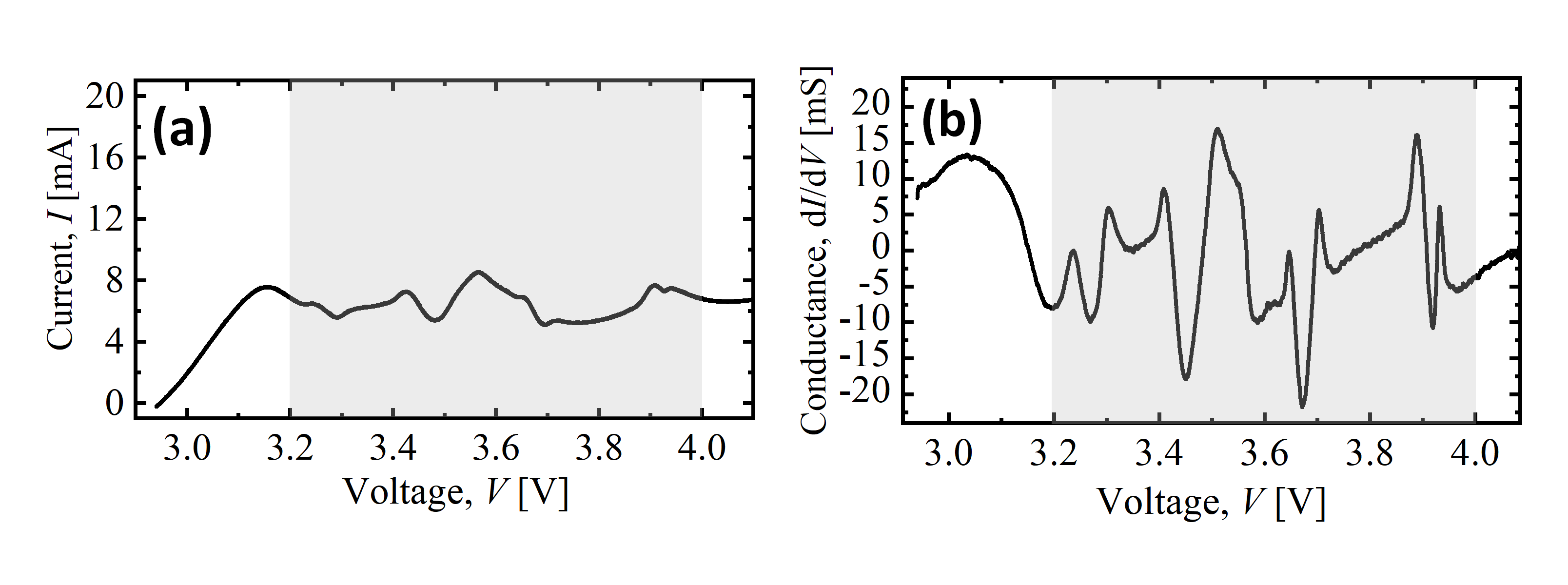}
\caption{{\bf Figure 3:} (a) Current, $I$ -- Voltage, $V$ characteristics derived from cyclic voltammetry experiment of the $K$ cation extraction (charging) process with \ce{K2Ni2TeO6} as the cathode in a two-electrode setup. $K$ metal was used as the counter electrode and the scanning rate was set to 0.1 mVs$^{-1}$.\cite{bard2001fundamentals} The current peaks at varied spread-out voltage values reflective of the small activation energy of $K$ in the honeycomb layered oxide material. (b) The conductance, $dI/dV \sim w\sigma$ of \ce{K2Ni2TeO6} during the charging process displaying conductance spikes at varied spread-out voltage values corresponding to extractions of a large number of K cations from multiple honeycomb lattice layers, $k \rightarrow k + w$, where the values of $w \in \mathbb{N} \gg 1$ and $\sigma$ cannot be separately determined from the results. Nonetheless, the sharp conductance spikes occur evenly distributed at a rough interval of 0.1 V within the voltage interval, 3.2 V to 4.0 V (gray/shaded region)}
\end{center}
\end{figure}

The low resolution is an artifact of the multi-layered nature of the materials, with each honeycomb lattice not only contributing active primitive cells with cationic sites during the extraction but also multiple sites getting activated at once at spread out external voltage values. Moreover, experimental data suggests that a large activation energy, $E_{\rm a} \gg E_{\rm a}^{\rm K} \simeq 121$ meV, where $E_{\rm a}^{\rm K}$ is the activation energy of potassium, K cations in the honeycomb lattice\cite{matsubara2020magnetism} and the presence of cationic vacancies before the extraction process would tend to disfavor the cation extraction process from occurring at evenly spread-out (low to high) voltage values during cycling, often leading to a solitary broad current peak centered at the high voltage regime, for instance, as can be seen in the $I$--$V$ cycling characteristics for $A_2$\ce{Ni2TeO6} with $A = \rm Li, Na, K$, since the activation energy for Li and Na is vastly greater than that of K, \textit{i.e.} $E_{\rm a}^{\rm Li} > E_{\rm a}^{\rm Na} > E_{\rm a}^{\rm K}$.\cite{kanyolo2021honeycomb}
Consequently, we have plotted only the Current, $I$ -- Voltage, $V$ characteristics of the extraction process for \ce{K2Ni2TeO6} in {\bf Figure 3}, which exhibits several distinct current peaks across varied voltage values.

Moreover, we can employ eq. (\ref{Gauss_Bonnet_eq}), which suggests the partition function, 
\begin{align}
    \mathcal{Z} \propto \sum_{\mathcal{A} \in h}\rho \left ( \Phi_{\rm AC}\right )\Bigr|_{\mathcal{A} \in h},
\end{align}
is given by the sum over different geometries of the manifold with distinct topology,
\begin{subequations}\label{QG_eq1}
\begin{align}
    \mathcal{Z} = \sum_{\mathcal{A} \in h}f_h\exp(2\pi k\,\chi(N, h)),\\
   \chi(N, h) = \frac{1}{4\pi}\int_{\mathcal{A} \in h} d^{\,2}x \sqrt{\det(g_{ab})}R(N),
\end{align}
\end{subequations}
where $R = R_{ab}g^{ab}$ is the 2D Ricci scalar, $R_{ab} = R_{acbd}g^{cd}$ is the 2D Ricci tensor and $R_{acbd} = K(g_{ab}g_{cd} - g_{ad}g_{bc})$ is the general form of the 2D Riemann tensor in Riemannian geometry with the equivalence, 
\begin{subequations}\label{QG_eq2}
\begin{align}
    \sum_{\mathcal{A} \in h} f_h \leftrightarrow \int \mathcal{D}[g_{ab}(\mathcal{A})],
\end{align}
assumed to be valid. Evidently, eq. (\ref{QG_eq1}) is the partition function of 2D quantum geometry in Euclidean signature\cite{gross1991two},
\begin{align}
    \mathcal{Z} = \int \mathcal{D}[g_{ab}]\exp\left (\frac{1}{2\kappa}\int_{\mathcal{A}} d^{\,2}x \sqrt{\det(g_{ab})}\,R\right ),
\end{align}
\end{subequations}
where the coupling constant corresponds to $\kappa = 1/k$. It is worth noting that, considering emergent geometries within crystals to describe defects is not entirely a novel idea, since it has been considered in great detail for disclinations and dislocations within the context of classical geometries with torsion.\cite{kleinert1987gravity, kleinert1988lattice, yajima2016finsler, holz1988geometry, verccin1990metric, kleinert2005emerging}

Finally, we are left to show that eq. (\ref{QG_eq1}) (and equivalently, eq. (\ref{QG_eq2})) obeys the necessary modular symmetries defined in eq. (\ref{modular_symmetry_eq}), which are imposed by the primitive cell of the cations in honeycomb layered materials. Nonetheless, eq. (\ref{modular_partition_eq}) approaches eq. (\ref{QG_eq1}) and eq. (\ref{QG_eq2}) in the limit $\kappa \rightarrow 0$, as required.

\subsection{Modular symmetries:}

The honeycomb lattice is spanned by the primitive basis, $dx$ and $dy$ defining a parallelogram enclosing $k = 1$ pair of cation sites. Since the unit cell in {\bf Figure 2} is a rhombus, we have $dx/dy = 1$.  Moreover, transforming the basis by the matrix,
\begin{subequations}
\begin{align}
J\begin{pmatrix}
dx \\ dy
\end{pmatrix}
= \begin{pmatrix}
dx' \\ dy'
\end{pmatrix},
\end{align}
where, 
\begin{align}
    J = \begin{pmatrix}
 \alpha & \beta \\
 \gamma & \delta \\
\end{pmatrix} \in \rm SL_2(\mathbb{Z}),
\end{align}
\end{subequations}
we find that $J_1$ and $J_2$, given in eq. (\ref{modular_symmetry_eq}), correspond to the re-scaling, $dx'/dy' = dx/dy + 1 = 2$ and the discrete rotation, $dx/dy \rightarrow dx'/dy' = -dy/dx = -1$, as illustrated in \textbf{Figure 4} and \textbf{Figure 5} respectively.  Moreover, we shall consider the modular form\cite{cohen2017modular} defined as, $g(dx, dy) = \int f(k)dk$ to completely characterize the honeycomb lattice in an invariant manner, under $J \in \rm SL_2(\mathbb{Z})$ with $dx/dy = k$. By definition, $g(dx', dy') = g(dx, dy)$ is invariant under the modular transformations, \textit{i.e.} $k \rightarrow J \cdot k = (\alpha k + \beta)/(\gamma k + \delta)$. Consequently, $f(k)$ transforms as a modular form of weight 2, 
\begin{align}
    f(J \cdot k) = (\gamma k + \delta)^2f(k).
\end{align}
Proceeding, we must take the large limit, $k \rightarrow \infty$, which spans the entire honeycomb lattice. Moreover, assuming $df(k)/dk = 0$, we obtain, $g(dx, dy) = kf(k)$. 

\begin{figure}
\begin{center}
\includegraphics[width=0.8\columnwidth,clip=true]{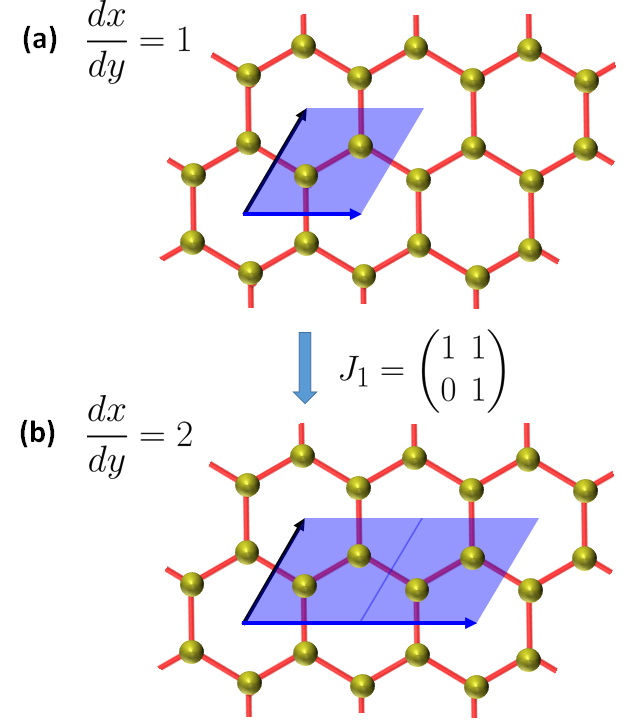}
 \caption{{\bf Figure 4:} The honeycomb lattice of cations depicting the action of the $J_1$ generator of $\rm SL_2(\mathbb{Z})$ on the primitive vectors. (a) The primitive vectors $dx$ and $dy$ of the primitive cell of the honeycomb lattice, where $dx/dy = 1$ is the number of pairs of cations enclosed within the primitive cell; (b) $J_1$ transformation corresponding to the re-scaling of the $dx$ primitive vector and hence an expansion of the unit cell, $dx/dy = 2$.}
 \end{center}
\end{figure}

\begin{figure}
\begin{center}
\includegraphics[width=\columnwidth,clip=true]{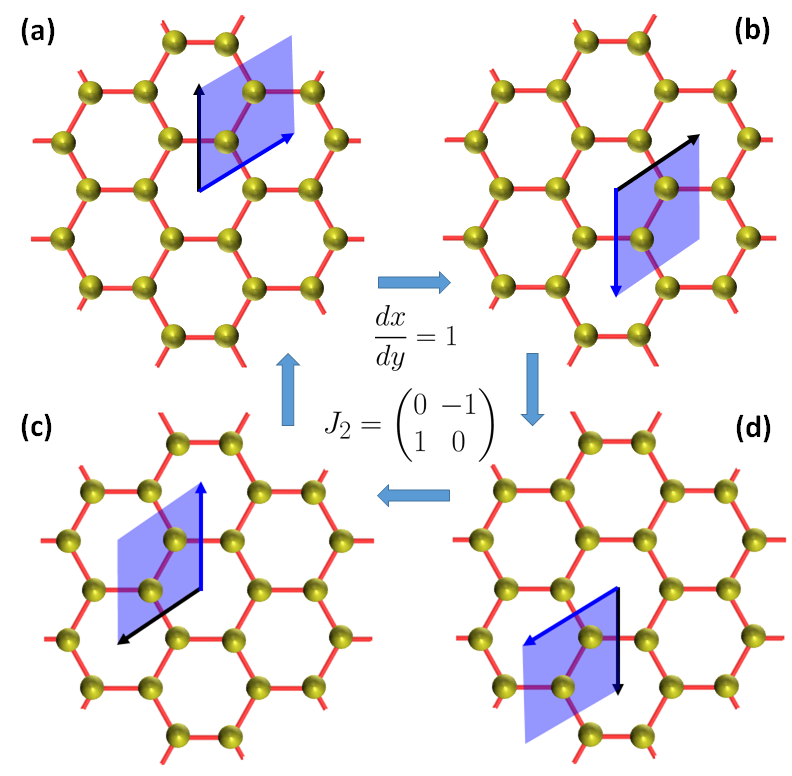}
 \caption{{\bf Figure 5:} The honeycomb lattice and discrete rotations ($dx/dy = 1$) generated by $J_2 \in \rm SL_2(\mathbb{Z})$ acting on the primitive cell. The primitive cell is rotated as shown in (a), (b), (c) and (d) by application of the $J_2$ transformation, such that $J_2^2 = -I_2$ corresponds to inversion of (a) to (d) and (b) to (c), where $I_2$ is the $2\times 2$ identity matrix, requiring that $J_2^4 = I_2$.}
\end{center}
\end{figure}

Now, consider the diffusion dynamics of the cations given in eq. (\ref{Liouville_eq}) and eq. (\ref{Langevin_eq}). Defining the velocity, $\vec{u} = \bar{\beta}\vec{p} = \exp(\Phi)d\vec{x}/ds$, where $\ell$ is the arc length interval along the proper length, $ds$, we obtain,
\begin{subequations}
\begin{multline}
   \frac{M}{2}\int_{\ell} ds = \frac{M}{2}\int_{\ell} \exp(-\Phi)\sqrt{dx^2 + dy^2}\\
   = \frac{M}{2}\int_{\partial \mathcal{A}} \vec{u}\cdot d\vec{x}
   = k\int_{\partial \mathcal{A}} \vec{p}\cdot d\vec{x}
   = k\int_{\partial\mathcal{A}} \vec{\nabla}\Phi_{\rm AC}\cdot d\vec{x}\\ = k\Phi_{\rm AC}\Bigr|_{\partial \mathcal{A}} = -2\pi k\chi,
\end{multline}
\end{subequations}
where we have used $\bar{\beta}M = N = 2k$ from eq. (\ref{k_eq}) and eq. (\ref{Gauss_Bonnet_eq}). Thus, setting $f(k) = -2\pi\chi(k)$ defines the modular form as the action of a particle of mass, $M/2$ in 2D Riemannian geometry, $g(dx, dy) = (M/2)\int ds$ or equivalently the exponent of eq. (\ref{QG_eq1}) (or eq. (\ref{QG_eq2})).

\subsection{Liouville conformal field theory:}

To elucidate further properties of the formalism, we shall consider the Liouville action\cite{alvarez2013random}, 
\begin{align}\label{Liouville_action_eq}
   S_{\omega} = \int \frac{d^{\,2}X}{\omega}\sqrt{\det(\tilde{g}_{ab})}\left (\tilde{g}^{ab}\frac{\partial\phi}{\partial X^{a}}\frac{\partial\phi}{\partial X^{b}} + K\exp(2\omega\phi) + Q(\omega)\tilde{R}\phi \right )
\end{align}
where 
$\tilde{g}_{ab} = \exp(-2\omega\phi)g_{ab}$ is the 2D metric tensor with $g_{ab}$ given in eq. (\ref{Liouville_eq}), $K$ is the Gaussian curvature associated with $g_{ab}$ and $\tilde{R}$ is the Ricci scalar associated with $\tilde{g}_{ab}$, $Q(\omega)$ is a parameter dependent on $k$ and genus $h$. Setting $\Phi = \omega\phi$ and $x^a = \omega X^a$, the metric in eq. (\ref{Liouville_action_eq}) reduces to the 2D identity matrix, $\tilde{g}_{ab} = \exp(-2\Phi)g_{ab} = \delta_{ab}$ requiring that the Ricci scalar vanishes, $\tilde{R} = 0$ getting rid of the last term even when $Q(\omega) \neq 0$. 

The Liouville action reduces to,
\begin{align}\label{Liouville_action_eq2}
   S_{\omega} = \omega\int d^{\,2}x\left (\vec{\nabla}\Phi\cdot\vec{\nabla}\Phi + K\sqrt{\det(g_{ab})} \right ), 
\end{align}
where $\sqrt{\det(g_{ab})} = \exp(2\Phi)$. Thus, for arbitrary $Q(\omega)$, eq. (\ref{Liouville_action_eq2}) can be varied with respect to $\Phi(\vec{x})$ to yield eq. (\ref{Liouville_eq}). Moreover, we define the path integral as, 
\begin{subequations}\label{path_int_eq}
\begin{align}
    \mathcal{Z} = \int\mathcal{D}[g_{ab},\Phi]\sum_j\exp\left (iS_{\omega_j}(\Phi, g_{ab}) \right ),
\end{align}
which, after summation over $j$ and functional integration over $\Phi$ yields, 
\begin{align}
    \mathcal{Z} = \int \mathcal{D}[g_{ab}]\cosh(2\pi k(\chi + \theta)),
\end{align}
\end{subequations}
where, 
\begin{subequations}
\begin{align}
    2\pi\theta = \frac{Area}{2}\int\frac{d^{\,2}p}{(2\pi)^2} \ln p^2 \rightarrow 2\pi\sum_p\ln p + C,
\end{align}
is the divergent vacuum energy of $\Phi$ with $g_{ab}$ and $\Phi$ approximated as separate non-interacting fields, $C$ is a constant that will be set to vanish by regularization, $Area = \frac{1}{2\pi}\int d^{\,2}x$ is the area element, $Area\int d^{\,2}p/(2\pi)^2 \rightarrow \sum_p$, $\vec{p}$ are the allowed momenta/energies of the bosonic field, $\Phi$, $\omega_j = (\omega, \bar{\omega})$, $\omega = ik$ and $\bar{\omega} = -ik$. Thus, Riemann zeta function regularization requires\cite{hawking1977zeta}, 
\begin{align}\label{regularization_eq}
    \theta(s) = \sum_p p^{-s}\ln p + C = \sum_p\frac{1}{p^{s - 1}},\\
    C = \sum_p\frac{1}{p^{s - 1}} - \sum_p p^{-s}\ln p,
\end{align}
\end{subequations}
where $\theta = \theta(s = 0)$. 

To elucidate the nature of eq. (\ref{path_int_eq}), we consider the Virasoro algebra\cite{polchinski1998string2},
\begin{subequations}\label{Virasoro_eq}
\begin{align}
    [L_n, L_m] = (n - m)L_{n + m} + \frac{c}{12}(n^3 - n)\delta_{(n + m), 0},\\
    [\bar{L}_n, \bar{L}_m] = (n - m)\bar{L}_{n + m} + \frac{\bar{c}}{12}(n^3 - n)\delta_{(n + m), 0},
\end{align}
\end{subequations}
spanned by two copies of commuting generators, $L_n = L_{-n}^{\dagger}$ and $\bar{L}_n = \bar{L}_{-n}^{\dagger}$ for all integers $n \in \mathbb{Z}$, where $[\bar{L}_n, L_m] = 0$, $c, \bar{c}$ is a real-valued constant (the central charge), satisfying $[L_n, c] = [L_n, \bar{c}] = [\bar{L}_n, c] = [\bar{L}_n, \bar{c}] = 0$ and $\delta_{m,n}$ is the Kronecker delta. Meanwhile, the representations are characterized by a highest weight primary state, $|L\rangle, |\bar{L}\rangle$ satisfying, $L_0|L \rangle = L|L \rangle$, $\langle L|L_0 = \langle L|L$ or $\bar{L}_0|\bar{L} \rangle = \bar{L}|\bar{L} \rangle$, $\langle \bar{L}|\bar{L}_0 = \langle \bar{L}|\bar{L}$, $L_{|n| \neq 0}|L \rangle = 0$, $\langle L|L_{-|n| \neq 0} = 0$ and $\bar{L}_{|n| \neq 0}|\bar{L} \rangle = 0$, $\langle \bar{L}|\bar{L}_{-|n| \neq 0} = 0$. The rest, $L_{-|n| \neq 0}|L \rangle \neq 0$, $\bar{L}_{-|n| \neq 0}|\bar{L} \rangle \neq 0$ and $\langle L|L_{|n| \neq 0} \neq 0$, $\langle \bar{L}|\bar{L}_{|n| \neq 0} = 0$ can be computed by applying the Virasoro algebra in eq. (\ref{Virasoro_eq}).

Proceeding, it is known that the field $V(\alpha) = \exp(2\alpha\phi)$ is primary when the conformal dimension is given by\cite{zamolodchikov1996conformal, nakayama2004liouville}, 
\begin{subequations}\label{marginal_eq}
\begin{align}
    L(k, h) = \alpha(k, h)(Q(\omega(k)) - \alpha(k, h)),\\
    \bar{L}(k, h) = -\alpha(k, h)(\bar{Q}(\bar{\omega}(k)) + \alpha(k, h)),
\end{align}
\end{subequations}
while the marginal condition for the primary field that guarantees conformal invariance of the theory is $L = \bar{L} = 1$ with $\alpha = \omega = -\bar{\omega}$ which yields, 
\begin{align}\label{Q_eq}
    Q(\omega) = \omega(k) + 1/\omega(k) = i(k - 1/k) = -\bar{Q}(\bar{\omega}),
\end{align}
with $\omega(k) = ik, \bar{\omega}(k) = -ik$ and $\bar{Q}(\bar{\omega}) = \bar{\omega} + 1/\bar{\omega}$. The central charge is given by, 
\begin{subequations}
\begin{align}
    c(\omega) = 1 + 6Q^2(\omega) = 1 + 6\bar{Q}^2(\bar{\omega}) = \bar{c}(\omega),
\end{align}
which can be written as, 
\begin{align}
    c(E, T) = 1 - 6\frac{(E - T)^2}{ET}, 
\end{align}
\end{subequations}
where $T = 1/\bar{\beta}$ is the temperature, $E = M/2$ is a defined energy in order for $k = E/T = \bar{\beta}M/2$ as required. The conformal dimension and spin are given by,  
\begin{subequations}\label{delta_eq}
\begin{align}
    \Delta = L + \bar{L} = 2\alpha(Q(\omega) - \alpha) = -2\alpha(\bar{Q}(\bar{\omega}) + \alpha),\\
    \sigma = i(L - \bar{L}) = 2\alpha i (Q(\omega) + \bar{Q}(\bar{\omega})) = 0,
\end{align}
\end{subequations}
respectively. However, the theory is known to be unitary only for $c = 1$ and $c = \infty$ corresponding to $k = 1$ and $k \rightarrow i\infty$ respectively. Since $k$ is considered real, we must have $k = 1$. Thus, $k \rightarrow \infty$ not only breaks discrete rotation symmetry but also breaks unitarity. Indeed, unlike eq. (\ref{Q_eq}), $Q \sim ik$ is not invariant under $k \rightarrow -1/k$, which breaks the discrete rotation symmetry generated by $J_2$, where $k = N/2$ is the number of primitive cells in the honeycomb lattice. 

Now, we consider the partition function, 
\begin{align}\label{partition_CFT_eq}
    \mathcal{Z} = \frac{1}{4}\sum_{j, h}f_h\langle L, \bar{L}|\left (q^{L_0 - c(h)/24}(\omega_j) \bar{q}^{\bar{L}_0 - \bar{c}(h)/24}(\bar{\omega}_j) \right )|L, \bar{L}\rangle = \sum_hf_h\cosh(2\pi\bar{\beta}E_{\rm CFT}),
\end{align}
where $q(\omega_j) = \exp(i2\pi\omega_j)$, $\bar{q}(\bar{\omega}_j) = \exp(-i2\pi\bar{\omega}_j)$ and $\omega_j = (\omega, \bar{\omega})$, $\bar{\omega}_j = (\bar{\omega}, \omega)$, arriving at the energy,
\begin{subequations}
\begin{align}
    E_{\rm CFT}(L, \bar{L}, c, \bar{c}) = \frac{M}{2}(L + \bar{L} - c/24 - \bar{c}/24),
\end{align}
with $k = \bar{\beta}M/2$ from eq. (\ref{k_eq}). Consequently,  
\begin{align}
    E_{\rm CFT}(k,h) = M\alpha(k, h)(Q(k) - \alpha(k, h)) - MQ^2(k)/4 - M/24 = \frac{M}{2}\left ( \chi(h) - \frac{1}{12} \right ),
\end{align}
where we have used eq. (\ref{delta_eq}) and defined the momentum of the primary field, $V(\alpha) = \exp(2\alpha\phi)$ as,
\begin{align}
    \alpha(k, h) = Q(k)/2 \pm i\sqrt{\chi(h)/2}.
\end{align}
\end{subequations}
Recall that the marginal condition is guaranteed by $L = \bar{L} = 1$, which now translates to, 
\begin{align}
   c = \bar{c} = 1 + 6Q^2(k) = 25 - 12\chi(h),
\end{align}
for which $\chi(h) = 2 - 2h$ yields, $c = \bar{c} = 1 + 24h$ or equivalently $Q^2(k) = -(k - 1/k)^2 = 4h$. 

In this case, unitarity is only achieved for $h = 0$, which corresponds to the two-sphere satisfying $k = 1$. Moreover, to reconcile with eq. (\ref{path_int_eq}) the extra factor of $-1/12$ must correspond to the vacuum energy, $\theta = \theta(s = 0)$ after regularization. This requires the momenta $p \in \mathbb{N}$ be positive integers, which yields the expression $\theta = 1 + 2 + 3 + \cdots = -1/12$ by regularization. Note that, for $k \geq 1$, all other positive values of $h$ cannot satisfy the marginal condition and hence must break conformal invariance.

\section{Discussion}

If the Euler characteristic of the $n$-th primitive cell can be associated with a manifold defined by the Poincar\'{e} polynomial,
\begin{align}\label{Betti_eq}
    P_n(Y) = b_n^{(0)} + b_n^{(1)}Y + b_n^{(2)}Y^2
\end{align}
where $b_n^{(p = 0, 1, 2)}$ is the $p$-th Betti number, the Euler characteristic of emergent manifold can be calculated as the Euler characteristic of the connected sum of emergent manifolds corresponding to the primitive cells, $\chi(\mathcal{A}_1\#\cdots\mathcal{A}_{n = h}) = b_n^{(0)} + b_n^{(2)} - \sum_{n = 1}^hb_n^{(1)}$ with topology, $h$, which can be decomposed into a Fourier series given by eq. (\ref{chi_eq}), where $q(k) = \exp(2\pi i k) = 1$ with $k \in \mathbb{N}$ and we have used the fact that only the $p = 1$-st Betti numbers are additive in a connected sum. Consequently, the simplest real-valued partition function that respects the modular symmetries, $J_1$ and $J_2$ in their respective limits of $k$ and is proportional to the partition function in eq. (\ref{QG_eq1}) (and hence eq. (\ref{QG_eq2})) corresponds to eq. (\ref{modular_partition_eq}), where the limit $h, k \rightarrow \infty$ breaks the discrete rotation symmetry, $J_2$ of the partition function, $\mathcal{Z}$ whilst promoting scale invariance, $J_1$. Mathematically, this is required since there exists no holomorphic modular forms of weight $2$, invariant under both $J_1, J_2 \in \rm SL_2(\mathbb{Z})$ transformations.\cite{cohen2017modular} 

Nonetheless, this can be remedied by considering the Euler characteristic proportional to the \textit{almost holomorphic} modular form of weight 2 in the large genus limit ($h \rightarrow \infty$)\cite{cohen2017modular}, 
\begin{align}
    \chi(h, \tau) = 2\left (E_2(h, \tau) - \frac{3}{\pi\Im(\tau)} \right ),
\end{align}
where $k \rightarrow k + i\epsilon = \tau$, 
\begin{align}
   E_2(h, \tau) = -24\sum_{n = 0}^h\sigma_1(n)q^n(\tau),
\end{align}
with $E_2(\tau, h \rightarrow \infty)$ the second Eisenstein series, $\sigma_1(0) = -1/24$ and $\sigma_1(n > 0)$ the sum of the divisors of the positive integer, $n$. Thus, the Betti numbers for the emergent geometry of the honeycomb lattice primitive cells correspond to the two-torus, 
\begin{subequations}
\begin{align}
    b_n^{(0)} = b_n^{(2)} = -24\sigma_1(0) = 1, b_n^{(1)} = 2,
\end{align}
where for such $h$ primitive cells in a connected sum, we set, 
\begin{align}
    \epsilon = \Im(\tau) = \frac{6/\pi}{2h - 48\sum_{n = 1}^h\sigma_1(n)q^n(k)}.
\end{align}
\end{subequations}
Thus, unlike in Liouville CFT, modular invariance and hence conformal invariance is guaranteed for all integer values of $h$ and $k$. However, it is broken for the primitive cell when $\chi_n \neq 0$, corresponding to a phase transition.\cite{kanyolo2022conformal}

To discuss the effects of such a phase transition, we note that, $J_2^2 = -I_2$, where,
\begin{align}
I_2 = \begin{pmatrix}
1 & 0\\ 
0 & 1
\end{pmatrix},
\end{align}
which implies the unit basis acquires a minus sign under $J_2^2$. Since $J_2$ exchanges the basis $dx$ with $dy$ and vice-versa, it corresponds to a discrete rotation when acting on the primitive cell. There are 4 such discrete rotations such that $J_2^{4n}$ correspond to complete $2\pi n$ rotations of the primitive cell, where $n \in \mathbb{N}$ is a real number. In addition, $J_2^{2(2n + 2)}$ exchanges one cationic site in the primitive cell with the other. Thus, under the exchange of two cations belonging to the same primitive cell, the transformation picks up a minus sign. This can be understood as the origin of the pseudo-degree of freedom we shall refer to as pseudo-spin, which distinguishes the two sub-lattices of the honeycomb lattice.\cite{mecklenburg2011spin, kanyolo2022conformal} 

Under specific conditions, the pseudo-spin of the graphene lattice can be linked to the spin of the electrons localized on the carbon atoms in the sub-lattice.\cite{mecklenburg2011spin} However, for cations in honeycomb layered oxides, no such identification can be affirmed. Nonetheless, the pseudo-spin degree of freedom, coupled with the $\rm SL_2(\mathbb{Z})$ group imply the partition function given in eq. (\ref{partition_eq}) that the underlying theory of cations is a conformal field theory whose ground state must avoid pseudo-spin frustration by pseudo-spin anti-ferromagnetic behavior, but nonetheless prevents the cations from forming a stable honeycomb lattice due to a repulsive exchange interaction which can be offset by pairing of opposite pseudo-spin degrees of freedom.\cite{kanyolo2022conformal} 

The effective theory for two pseudo-spin cations ($j = 1, 2$) in $k = \beta M/2 = N/2$ non-interacting honeycomb primitive cells corresponds to the 1D Ising Hamiltonian\cite{baxter1982inversion}, 
\begin{align}
    \mathcal{H}_{\rm Ising} = - \frac{1}{2}\sum_{j,j' = 1,2} A_{jj'}(h)\sigma_z^j\sigma_z^{j'} - B(h)\sum_{j = 1,2}\sigma_z^j,
\end{align}
where, 
\begin{align}
    A_{jj'}(h) = \begin{pmatrix}
 0 & A(h) \\
 A(h) & 0 \\
 \end{pmatrix},
\end{align}
$B(h) = 2\pi M\chi(h)/2$ is the pseudo-magnetic field\cite{georgi2017tuning, kanyolo2022conformal} in the $z$-direction interacting with the pseudo-spins, $\sigma_z$ which is taken to be proportional to the Euler characteristic, $\chi(h)$, while $A(h)$ is the Heisenberg term representing the exchange interaction, assumed to depend on the genus, $h = \nu + 1$ with $\nu$ the cationic vacancy number.

This Ising model is exactly solvable, where standard calculation for the partition function yields\cite{grosso2013solid}, 
\begin{align}
    \mathcal{Z} = {\rm Tr}_{h,\sigma}\exp(-\bar{\beta}\mathcal{H}_{\rm Ising}) = {\rm Tr}_{h,\sigma} P = \sum_h\lambda_+(h) + \sum_h\lambda_-(h) = \sum_h f_h\exp(2\pi k\chi(h)),
\end{align}
where ${\rm Tr}_{h, s}$ is the trace over the genus $h$ and spins $\sigma$, and $\lambda_{\pm}$ are the eigenvalues of the transfer matrix,
\begin{align}
P = \begin{pmatrix}
\exp(-\bar{\beta}E_{\uparrow}) & \exp(-\bar{\beta}E_{\uparrow\downarrow})\\
 \exp(-\bar{\beta}E_{\downarrow\uparrow}) & \exp(-\bar{\beta}E_{\downarrow})\\
\end{pmatrix},
\end{align}
given by, 
\begin{align}
    \lambda_{\pm} = \exp(\bar{\beta}A)\cosh\bar{\beta}B\pm \sqrt{\exp(2\bar{\beta}A)\sinh^2(\bar{\beta}B) + \exp(-2\bar{\beta}A)},
\end{align}
with,
\begin{align}
   E_{\uparrow\downarrow} = E_{\downarrow\uparrow} = A,\,\, E_{\uparrow} = -(B + A),\,\, E_{\downarrow} = (B - A).
\end{align}
Thus, the non-interacting system of $k$ primitive cells occupied by pairs of $N = 2k$ cations interacting via their pseudo-spins and the pseudo-magnetic field yields the partition function in eq. (\ref{partition_eq}) where $f_h = 2\exp(\bar{\beta}A(h))$, which takes on varied values for different topology configurations, $h$. Moreover, the exponents of components of the transfer matrix correspond to the pseudo-spin energy states, where $E_{\uparrow} - E_{\downarrow} = 2B$ corresponds to a gapped phase where the honeycomb lattice bifurcates into bilayers with energies $E_{\uparrow}$ and $E_{\uparrow}$ due to a finite pseudo-magnetic field, $B \neq 0$, whereas $E_{\uparrow\downarrow} = E_{\downarrow\uparrow} = A$ correspond to the ferromagnetic ($A > 0$) and anti-ferromagnetic ($A < 0$) alignment of the pseudo-spins.

To avoid pseudo-spin frustration when $B = 2\pi M\chi(h)/2 = 0$ (two-torus, $\chi(h) = 0$), the honeycomb lattice must be anti-ferromagnetic described by the singlet bound state, $(|\uparrow\downarrow\rangle - |\uparrow\downarrow\rangle)/\sqrt{2}$, ($\langle\sigma_z^1\sigma_z^2\rangle = -3/4, \langle \sum_j\sigma_z^j\rangle = 0$), with $A = -|A| < 1$, which disfavors the ferromagnetic condition.\cite{blundell2003magnetism} Nonetheless, for a finite pseudo-magnetic field, $B \neq 0$ ($\chi(h) \neq 0$)  the triplet bound state, $(|\uparrow\downarrow\rangle + |\uparrow\downarrow\rangle)/\sqrt{2}$, $\langle\sigma_z^1\sigma_z^2\rangle = 1/4, \langle \sum_j \sigma_z^j \rangle = 1$) is allowed, corresponding to other topology configurations. We are interested in $\chi(h) = 2$, corresponding to the unitarity condition for marginal fields in Liouville conformal field theory. In this case,
\begin{subequations}
\begin{align}
    \mathcal{H}_{\rm Ising} = -2A\left \langle \sigma_z^1\sigma_z^2\right \rangle - B\sum_j \left \langle \sigma_z^j \right \rangle \equiv -\frac{1}{2}\int_{S^2} \left (J_{\rm RKKY} + MK\right ) = \int_{S^2}\rho_{\rm 2D}(r),
\end{align}
is the potential energy of the resultant bond, $\vec{r} = \vec{r}_{\uparrow} - \vec{r}_{\downarrow}$ is the 3D relative position of the pseudo-spin up and down cations, $x = |\vec{r}|$, $K = 1/r^2$ is the Gaussian curvature of the two-sphere ($S^2$) and due to the point like nature of the Fermi surface, $J_{\rm RKKY}(r) \equiv -2Mr_0/3r^3$ is taken to be the non-oscillatory Ruderman-Kittel-Kasuya-Yosida (indirect exchange) interaction mediated by the conduction electrons of the cations, where $r_0 \geq 0$ is a distance scale to be determined.\cite{brey2007diluted, cao2019rkky} Thus, the energy density (integrand),
\begin{align}
   \rho_{\rm 2D}(r) = -J_{\rm RKKY}(r)/2 - MK(r)/2 = M\left (\frac{r_0}{3r^3} - \frac{1}{2r^2} \right ),
\end{align}
\end{subequations}
corresponds to a metallophilic interaction\cite{kanyolo2022conformal} between the pseudo-spins in the primitive cell separated by a distance, $r$ apart in a stable bond forming bilayers. The bilayers are stable when $dV(r)/dr = 0$ and $d^2V(r)/dr^2|_{r = r_0} > 0$, corresponding to the separation distance $r = r_0$ between the honeycomb sub-lattices, which can be determined experimentally.\cite{masese2021honeycomb, kanyolo2022conformal} Consequently, this finite distance scale breaks scale/conformal invariance of the theory.

In conclusion, we have constructed a consistent framework to treat cationic vacancies in honeycomb layered materials as topological defects, $h$, by relating the Euler characteristic of the manifold to modular symmetries and 2D quantum geometries.\cite{gross1991two} The framework predicts integer conductance spikes during (de-)intercalation process, proportional to the number of active cation sites, $k \rightarrow k + w$ participating in the diffusion process at high resolution, $w \sim 1 \in \mathbb{N}$, which remain unobserved. Nonetheless, the framework greatly elucidates the geometric nature of the diffusion process which occur in these novel materials, and the crucial role played by cationic vacancies as topological defects, and hence should find great utility in finding avenues for performance optimization of such cathode materials for energy storage.\cite{masese2021topologicalK2Ni2TeO6, masese2021mixed} Further theoretical, computational and experimental treatments and applications are beyond the scope of the present work.\cite{kanyolo2021partition, kanyolo2021honeycomb, kanyolo2022conformal} 

\begin{addendum}
 
\item[Acknowledgments]
The authors acknowledge the financial support of TEPCO Memorial Foundation, Japan Society for the Promotion of Science (JSPS KAKENHI Grant Numbers 19K15685 and 21K14730) and Japan Prize Foundation. The authors also acknowledge fruitful discussions with D. Ntara during the cradle of the ideas herein, and especially the rigorous proofreading work on the manuscript done by Edfluent. Both authors are grateful for the unwavering support from their family members (T. M.: Ishii Family, Sakaguchi Family and Masese Family; G. M. K.: Ngumbi Family). 
 
\item[Competing Interests] 
The authors declare that they have no competing financial interests.
 
\item[Correspondence] 
Correspondence and requests for materials and/or clarification of aspects related to the model should be addressed to any or both of the authors: Titus Masese, PhD: titus.masese@aist.go.jp and Godwill Mbiti Kanyolo, PhD: gmkanyolo@mail.uec.jp; gmkanyolo@gmail.com.

\end{addendum}

\end{document}